\documentclass[aps,prd,nofootinbib,twocolumn,preprintnumbers]{revtex4}
\usepackage{bbold}
\usepackage{eufrak}

\usepackage{graphicx}
\usepackage{wasysym}
\newcommand{\bm}[1]{\mbox{\boldmath $#1$}}

\begin{document}
\title{Tensors, non-Gaussianities, and the future of potential reconstruction}
\author{Brian A. Powell} \email{brian.powell@ipmu.jp}
\affiliation{Institute for the Physics and Mathematics of the Universe,
University of Tokyo, Kashiwa, Chiba 277-8568, Japan}
\author{Konstantinos Tzirakis} \email{ct38@buffalo.edu}
\author{William H. Kinney} \email{whkinney@buffalo.edu}

\affiliation{Dept. of Physics, University at Buffalo,
        The State University of New York, Buffalo, NY 14260-1500}
\date{\today}

\preprint{IPMU 08-0114}
\begin{abstract}
\noindent
We present projections for reconstruction of the inflationary potential expected from ESA's
upcoming Planck Surveyor CMB mission.  We focus on the effects that 
tensor perturbations and the presence of non-Gaussianities have on
reconstruction efforts in the context of non-canonical inflation models. We consider potential constraints for different combinations of
detection/null-detection of tensors and non-Gaussianities.  We perform
Markov Chain Monte Carlo and flow analyses on a simulated
Planck-precision data set to obtain constraints.  We find that a failure
to detect non-Gaussianities  precludes a successful
inversion of the primordial power spectrum, greatly affecting
uncertainties, even in the
presence of a tensor detection.  In the absence of a tensor detection,
while unable to determine the energy scale of inflation, an observable
level of non-Gaussianities provides correlations between the errors of
the potential parameters, suggesting that constraints might be improved
for suitable combinations of parameters.  Constraints are optimized for a
positive detection of both tensors and non-Gaussianities.
\end{abstract}

\maketitle

\section{Introduction}
A faithful reconstruction of the inflaton potential is paramount to
understanding the epoch of inflation.  Recent cosmological observations
\cite{Hinshaw:2008kr,Gold:2008kp,Nolta:2008ih,Dunkley:2008ie,Komatsu:2008hk}
have revealed much about the  
possible forms of the potential
\cite{Kinney:2006qm,Lesgourgues:2007gp,Peiris:2008be,Lorenz:2008je}.  Future CMB missions, such as ESA's Planck
Surveyor \cite{planck}, and ground-based polarization experiments such as BICEP \cite{Yoon:2006jc}, promise to
further hone our understanding of the early universe.  The Planck Surveyor will
provide improved measurements of the temperature and E-mode polarization
anisotropies of the CMB out to $\ell = 3000$, and will detect the B-mode
polarization characteristic of gravity waves if the tensor/scalar ratio, $r$, is
greater than around  0.05.  Planck will also provide the first high-quality measurement of any
departure from Gaussianity exhibited by the CMB temperature fluctuations, with a
projected sensitivity $|f_{NL}| \gtrsim 5$ \cite{Komatsu:2002db,Creminelli:2003iq}.  

Recent times have also seen exciting progress in inflationary model building,
and much is now understood about how to implement inflation within string
theory (see
\cite{Cline:2006hu,Kallosh:2007ig,Burgess:2007pz,McAllister:2007bg} for
reviews).  One particularly exciting realization 
is provided by the dynamics of D-branes evolving in a higher-dimensional, warped
spacetime.  A novel aspect of these models is the existence of a speed limit on
the field space, resulting from causality restrictions on the motion of the
branes in the bulk spacetime.   This allows for slow roll to be achieved even in
the presence of a steep potential.  In terms of effective field theory, the speed
limit is enforced by non-canonical kinetic terms in the
DBI action describing the motion of the D-branes.  These models, termed DBI
inflation \cite{Silverstein:2003hf}, are further distinguished by the fact that scalar perturbations
propagate at a sound speed $c_s \leq 1$.  In the limit $c_s \ll 1$, these
fluctuations exhibit strong non-Gaussianities
\cite{Alishahiha:2004eh,Chen:2006nt,Spalinski:2007qy,Bean:2007eh,LoVerde:2007ri}.  Non-Gaussianities of this type were first popularized in the
context of {\it k-inflation} \cite{ArmendarizPicon:1999rj}, a model-independent construction in which higher-derivative kinetic terms drive
inflation in the absence of any potential energy.   This phenomenon is
a general characteristic of models with non-canonical kinetic terms in the
Lagrangian. 

With the advent of these more sophisticated theoretical constructions, and the increasingly
accurate cosmological probes planned to study them, more generalized methods of inflationary
reconstruction have been developed.  Bean {\it et al.} \cite{Bean:2008ga} present a formalism for
reconstructing a general action -- both potential and non-canonical kinetic energy functions
can be linked to observations.  The new functional degree of freedom presented by the
non-canonical form of the kinetic energy typically gives rise to additional observational
signatures.  For example, the speed of sound, $c_s$, becomes an additional dynamical
degree of freedom in k-inflation and DBI, giving rise to non-Gaussian density
fluctuations.
However, $c_s$ also affects the form of the primordial power spectra.  A successful
reconstruction of the inflaton potential, $V(\phi)$, thus requires a measurement of
$c_s$, since otherwise the expressions giving the spectral parameters cannot be uniquely
inverted to obtain $V(\phi)$.  An accurate reconstruction must also include a detection
of the tensor power spectrum.  Currently, neither of these quantities have been measured:
present bounds on any non-Gaussianities lie within $-9 < f_{NL}^{\rm local} < 111$ and
$-151 < f_{NL}^{\rm equil} < 253$ (95\% CL) \cite{Komatsu:2008hk}, and $r < 0.2$ (95\%
CL) with WMAP+BAO+SN \cite{Komatsu:2008hk} and $r<0.35 $ (95\% CL) for WMAP+ACBAR
\cite{Kinney:2008wy} for power-law
spectra, but might be
larger if more exotic spectra are considered \cite{Powell:2007gu}. 

In this paper, we make reconstruction projections for the Planck satellite
within the context of this larger inflationary parameter space, namely, one that contains
$c_s$ as an additional degree of freedom.  We seek to determine what combination of observations will be
required to most successfully reconstruct the potential.  In the best of all worlds,
future observations will detect both tensors and non-Gaussianities, telling us not only
that exotic physics is responsible for inflation, but also allowing for a successful
reconstruction program.  In this paper, we focus particularly on cases in which only one of these
observables is positively detected.  

In Section 2, we review potential reconstruction for the case of non-canonical inflation.
In Section 3, we perform a Bayesian analysis on a simulated
Planck-precision data-set for which there is a positive tensor detection, $r = 0.1$, but
no detection of non-Gaussianities, $|f_{NL}| < 5$.  This corresponds to a universe in
which canonical inflation took place (with $c_s = 1$), or, for example, DBI inflation
with $c_s \in [0.25,1]$.  We obtain constraints on $V(\phi_0)$,
$V'(\phi_0)$, and $V''(\phi_0)$ for both possibilities and find that the error bars are
significantly larger if $c_s$ is taken as a free parameter within the above range.
Our inability to resolve $c_s$
prohibits us from using the shape of the power spectra alone to reconstruct $V(\phi)$. 
In Section 4, we investigate this problem using inflationary flow methods.  The flow method is a natural
extension of the MCMC analysis, allowing higher dimensional parameter spaces to be explored with relative
ease. We reproduce
the results obtained in Section 3, and then investigate the effect of allowing the speed of sound to vary
over the course of inflation.  We find that constraints on $V_0''$ loosen considerably as this new degree
of freedom is introduced.  Next, we use flow methods to investigate the observational possibility of a
positive detection of $f_{NL}$ and a null detection of $r$. We find that
constraints on the individual potential parameters are worse in this case
than in the case where tensors are detected, but that the errors are
highly correlated, and tight bounds can be placed on combinations of
these parameters. Finally, we consider the case for which {\it both} $r$
and $f_{NL}$ are measured. This is clearly the best case scenario for near future experiments, since it will not only place bounds on the potential parameters from above and below, but it will also be a smoking gun for exotic physics.  
In Section 5 we present conclusions.

\section{Inflation models with an arbitrary speed of sound}
\subsection{Flow parameters and reconstruction}
In this section we consider the case of a scalar field with a general Lagrangian $\mathcal{L}(X,\phi)$, where $\phi$ is the field that is responsible for inflation ({\it the inflaton}) and $X=(1/2)g^{\mu \nu}{\partial_\mu \phi}{\partial_\nu \phi}$. If the inflaton field is coupled to Einstein gravity, the total action $S$ can be written \cite{Garriga:1999vw}
\begin{equation}
\label{eq:action}
S=\int d^4x\sqrt{-g}\left[\frac{1}{2}M_{\rm Pl}^2R+\mathcal{L}(X,\phi)\right],
\end{equation}
where $R$ is the curvature scalar and we work in units of the
reduced Planck mass, $M_{\rm Pl} = m_{\rm Pl}/\sqrt{8\pi}$.
We can also describe the inflaton field $\phi$ as a perfect fluid by expressing the energy-momentun tensor $T^{\hspace{0.15cm} \mu}_\nu$ in the following way
\begin{equation}
\label{eq:perfect fluid}
T^{\hspace{0.15cm} \mu}_\nu=(\rho+P)u_{\nu}u^{\mu}-P\delta^{\hspace{0.15cm} \mu}_\nu,
\end{equation}
where the pressure $P$ is identified with the Lagrangian $\mathcal{L}(X,\phi)$, $\rho$ is the energy density of the field,
\begin{equation}
\label{eq:energy density}
\rho=2X\frac{\partial \mathcal{L}(X,\phi)}{\partial X}-\mathcal{L}(X,\phi)=2X\mathcal{L},_{X}-\mathcal{L},
\end{equation}
and
\begin{equation}
\label{eq:four velocity}
u_{\nu}=\frac{\partial_{\nu}\phi}{\sqrt{2X}},
\end{equation}
is the fluid 4-velocity.
Assuming that the four-dimensional metric is of the Friedmann-Robertson-Walker (FRW) form
\begin{equation}
\label{eq:FRW}
g_{\mu \nu} = {\rm diag}\left(-1,a^2(t),a^2(t),a^2(t)\right),
\end{equation}
the Friedmann and the
acceleration equations are \cite{Bean:2008ga}
\begin{eqnarray}
\label{eq:Friedmann}
H^2 \equiv \left(\frac{\dot a}{a}\right)^2&=&\frac{1}{3M_{\rm Pl}^2}\rho, \cr \frac{\ddot a}{a}&=&-\frac{1}{6M_{\rm Pl}^2}(\rho+3P).
\end{eqnarray}
The speed at which 
fluctuations propagate relative to the homogeneous
background is determined by the sound speed of the fluid,
\begin{eqnarray}
\label{eq:speed of sound}
c_{s}^2 \equiv \frac{dP}{d\rho}=\frac{P,_{X}}{P,_{X}+2XP,_{XX}}=\frac{1}{\gamma^2}.
\end{eqnarray}
Inflation models driven by canonical scalar fields satisfy
\begin{eqnarray}
\label{eq:P,XX for canonical}
P,_{XX}=0,
\end{eqnarray}
and so fluctuations travel at the speed of light
\begin{equation}
\label{eq:speed of sound for canonical}
c^2 =1.
\end{equation}
However, in general non-canonical models, fluctuations can propagate at
speeds much lower, resulting in non-Gaussian adiabatic fluctuations
\cite{Alishahiha:2004eh,Chen:2006nt,Spalinski:2007qy,Bean:2007eh}.
In terms of the comoving curvature perturbation, $\zeta$, the
existence of non-Gaussianities is reflected in a non-trivial
three-point function, $\langle \zeta_{{\bf k}_1}\zeta_{{\bf
k}_2}\zeta_{{\bf k}_3}\rangle$.
In non-canonical models the dominant source of non-Gaussianity is due to
configurations for which the wavenumbers $k_1 = k_2 = k_3$, forming equilateral triangles in
Fourier space.  To lowest order in slow roll, the magnitude of this
non-Guassianity is given by the parameter \cite{Chen:2006nt},  
\begin{equation}
\label{ng}
f_{NL}^{\rm equil} = \frac{35}{108}\left(\frac{1}{c_s^2} - 1\right) -
\frac{5}{81}\left(\frac{1}{c_s^2}-1-2\Lambda\right),
\end{equation}
with
\begin{equation}
\Lambda = \frac{X^2 P,_{XX} + \frac{2}{3}X^3P,_{XXX}}{XP,_X + 2X^2P,_{XX}}.
\end{equation}

In the context of more general actions such as Eq. (\ref{eq:action}), the
process of reconstructing the physics of inflation requires an
understanding of both the potential energy of the field, $V(\phi)$, and
the kinetic term.  
Such a formalism was recently developed in Ref.
\cite{Bean:2008ga} in terms of different hierarchies of {\it flow parameters}.  Such
parameters naturally arise in the Taylor expansions of the functions $H$, $\gamma$,
and $\mathcal{L},_{X}$, which together completely specify the inflationary
dynamics.  In this paper, we fix the gauge by setting \cite{Agarwal:2008ah}
\begin{equation}
\label{eq:lagrangian equals gamma}
\mathcal{L},_{X}=c_{s}^{-1},
\end{equation}
which leads to the following hierarchies of flow parameters
\cite{Peiris:2007gz},
\begin{eqnarray}
\label{eq:flowparams}
\epsilon\left(\phi\right) &\equiv& \frac{2 M_{\rm Pl}^2}{\gamma\left(\phi\right)} \left(\frac{H'\left(\phi\right)}{H\left(\phi\right)}\right)^2,\cr
\eta\left(\phi\right) &\equiv& \frac{2 M_{\rm Pl}^2}{\gamma\left(\phi\right)} \frac{H''\left(\phi\right)}{H\left(\phi\right)},\cr &\vdots& \cr
{}^\ell \lambda\left(\phi\right) &\equiv& \left(\frac{2 M_{\rm Pl}^2}{\gamma\left(\phi\right)}\right)^{\ell} \left(\frac{H'\left(\phi\right)}{H\left(\phi\right)}\right)^{\ell - 1} \frac{1}{H\left(\phi\right)} \frac{d^{\ell + 1} H\left(\phi\right)}{d \phi^{\ell + 1}},\cr
s\left(\phi\right) &\equiv& \frac{2 M_{\rm Pl}^2}{\gamma\left(\phi\right)} \frac{H'\left(\phi\right)}{H\left(\phi\right)} \frac{\gamma'\left(\phi\right)}{\gamma\left(\phi\right)},\cr
\rho\left(\phi\right) &\equiv& \frac{2 M_{\rm Pl}^2}{\gamma\left(\phi\right)}  \frac{\gamma''\left(\phi\right)}{\gamma\left(\phi\right)},\cr &\vdots& \cr
{}^\ell \alpha\left(\phi\right) &\equiv& \left(\frac{2 M_{\rm Pl}^2}{\gamma\left(\phi\right)}\right)^{\ell} \left(\frac{H'\left(\phi\right)}{H\left(\phi\right)}\right)^{\ell - 1} \frac{1}{\gamma\left(\phi\right)} \frac{d^{\ell + 1} \gamma\left(\phi\right)}{d \phi^{\ell + 1}},
\end{eqnarray}
where $\ell = 2,\cdots,\infty$ is an integer index. For the remainder of
this paper, we will refer to the parameters $^{\ell}\lambda(\phi)$ and
$^{\ell}\alpha(\phi)$ as the $\ell$-th components of the {\it $H$-tower}
and {\it $\gamma$-tower} respectively. In the case where $\gamma=1$, the
$\gamma$-tower vanishes and the above hierarchy reduces to the flow
hierarchy of a canonical scalar field \cite{Liddle:1994dx,Kinney:2002qn}.

We next Taylor expand the Hubble parameter about $\phi_0 = 0$,
\begin{equation}
\label{eq:polynomial H}
H(\phi)=H_{0}\left[1+\displaystyle\sum_{\ell=1}^{\mathcal{M}+1} A_{\ell}\left(\frac{\phi}{M_{\rm Pl}}\right)^{\ell}\right],
\end{equation}
and  the inverse speed of sound
\begin{equation}
\label{eq:polynomial gamma}
\gamma(\phi)=\gamma_{0}\left[1+\displaystyle\sum_{\ell=1}^{\mathcal{N}+1} B_{\ell}\left(\frac{\phi}{M_{\rm Pl}}\right)^{\ell}\right],
\end{equation}
where the expansions are truncated at orders $\mathcal{M}+1$ and $\mathcal{N}+ 1$, respectively,
\begin{eqnarray}
\label{eq:Liddle truncation}
\frac{d^{\ell}H(\phi)}{d\phi^{\ell}}&=&0 \hspace{0.5cm} {\rm for} \hspace{0.5cm} \ell \geq \mathcal{M}+2, \cr \frac{d^{\ell}\gamma(\phi)}{d\phi^{\ell}}&=&0 \hspace{0.5cm} {\rm for} \hspace{0.5cm} \ell \geq \mathcal{N}+2. 
\end{eqnarray}
In fact, this truncation is equivalent to simply setting
$^{\mathcal{M}+1}\lambda =\,\, ^{\mathcal{N}+1}\alpha=0$ in the flow hierarchy, for reasons that will
become clear in Section IV. 

The coefficients $A_{\ell}$ and $B_{\ell}$ are given by
\cite{Peiris:2007gz}
\begin{eqnarray}
\label{Als}
A_{1}&=&\sqrt{\epsilon_{0}\gamma_{0}/2},\cr A_{\ell+1}&=& \frac{(\gamma_{0}/2)^{\ell}}{(\ell+1)!A_{1}^{\ell-1}}\left(^\ell\lambda_{0}\right), 
\end{eqnarray}
and
\begin{eqnarray}
\label{Bls}
B_{1}&=&\frac{s_{0}\gamma_{0}/2}{A_{1}},\cr B_{\ell+1}&=& \frac{(\gamma_{0}/2)^{\ell}}{(\ell+1)!A_{1}^{\ell-1}}\left(^{\ell}\alpha_{0}\right).
\end{eqnarray}
While this particular parameterization restricts $H(\phi)$ and
$\gamma(\phi)$ to be polynomials, these functions can be made more
general by truncating Eq. (\ref{eq:flowparams}) at arbitrarily high order. 
The values of the flow parameters at some $\phi=\phi_{0}$ fully define the functions $H(\phi)$ and $\gamma(\phi)$, completely determining
the theory in the gauge $\mathcal{L},_{X} = c_s^{-1}$.  

\subsection{DBI Inflation}
In what follows, we will study DBI inflation as a prototype non-canonical model.
The only model-dependent results in this paper are those relating non-Gaussianities to
the sound speed, $c_s$.  However, as we will discuss further, the qualitative nature
of our results should apply generically to non-canonical models. 
DBI inflation refers to models arising from the action of D-branes in flux 
compactifications of type-IIB string theory \cite{Kachru:2003aw,Kachru:2003sx,Silverstein:2003hf}.  Typically, the inflaton field
parameterizes the separation of a probe D3-brane and a stack of $\bar{\rm D3}$-branes in
a warped throat geometry.  This geometry is defined by the line element
\cite{Douglas:2006es},
\begin{equation}
{\rm d}s_{10}^2 = h^{-1/2}(y)g_{\mu \nu}{\rm d}x^\mu{\rm d}x^\nu + h^{1/2}(y)\left({\rm
d}\rho^2 + \rho^2{\rm d}s^2_{X_5}\right),
\end{equation}
where the internal space is a cone over the five-manifold $X_5$.  
The inflaton field is proportional to the throat coordinate $\rho$ by $\phi =
\sqrt{T_3}\rho$, where $T_3$ is the tension on the D3-brane. The Lagrangian for the
inflaton is then of the form \cite{Kachru:2003sx} 
\begin{equation}
\label{ldbi}
\mathcal{L} = -f^{-1}(\phi)\sqrt{1+f(\phi)g^{\mu\nu}\partial_\mu \phi \partial_\nu
\phi} + f^{-1}(\phi) - V(\phi),
\end{equation}
where $V(\phi)$ is the scalar field potential and the function $f(\phi)$ is related to the
warp factor $h(\phi)$ by
\begin{equation}
f(\phi) = \frac{1}{T_3h^4(\phi)}.
\end{equation}
The $\gamma$-factor defined by 
\begin{equation}
\label{eq:gamma}
\gamma = \frac{1}{\sqrt{1-f(\phi)\dot{\phi}^2}},
\end{equation}
enforces a speed limit on the moduli space,
arising from causality constraints on the motion of the probe brane in the throat.
This phenomenon mitigates the $\eta$-problem associated with potentials
derived from supergravity, allowing for otherwise too steep potentials to
successfully drive inflation.

The equations of motion are easily obtained from our previous results. The Friedmann equation (\ref{eq:Friedmann}) becomes \cite{Spalinski:2007kt},
\begin{equation}
\label{eq:Hamjacobi for DBI 1}
3M_{\rm Pl}^2H^2(\phi)-V(\phi) = \frac{\gamma(\phi)-1}{f(\phi)},
\end{equation}
which gives 
\begin{equation}
\label{me}
\dot \phi=-\frac{2M_{\rm Pl}^2}{\gamma(\phi)}H'(\phi).
\end{equation}
Equation (\ref{eq:gamma}) then becomes,
\begin{equation}
\label{eq: gamma H'}
\gamma=\sqrt{1+4M_{\rm Pl}^4f(\phi)\left[H'(\phi)\right]^2}.
\end{equation}
These results lead to the Hamilton-Jacobi equation,
\begin{equation}
V(\phi)=3M_{\rm Pl}^2H^2(\phi)-4M_{\rm Pl}^4\frac{H'^2(\phi)}{\gamma(\phi)+1},
\end{equation}
which can be used to reconstruct the scalar potential given the functions $H(\phi)$
and $\gamma(\phi)$.  The warp factor of the geometry can also be reconstructed from
these two functions via Eq. (\ref{eq: gamma H'}).  The Taylor expansions 
Eq. (\ref{eq:polynomial H}) and (\ref{eq:polynomial gamma}) can then be used to obtain
exact expressions for $V(\phi_0)$ and its derivatives.  In this work, we will only
require expressions for the first two derivatives of $V(\phi)$.  After some tedious
but straightforward algebra we obtain,
\begin{eqnarray}
\label{V V' V''}
V(\phi_{0})&=&M_{\rm Pl}^2H^2\left(3-2\epsilon\Gamma\right), \cr \nonumber\\ V'(\phi_{0})&=&M_{\rm Pl}H^2\sqrt{2\epsilon\gamma}\left(3-2\eta\Gamma+s\Gamma^2\right), \cr \nonumber\\ V''(\phi_{0})&=& H^2\gamma\left[3\left(\epsilon+\eta\right)-2\left(\eta^2+\xi\right)\Gamma \right. \cr &&\left. +2s\left(2\eta-s\Gamma\right)\Gamma^2+\epsilon\rho\Gamma^2\right],
\end{eqnarray}
where
\begin{equation}
\Gamma = \frac{\gamma}{\gamma +1}.
\end{equation}
In the next section, we utilize these results to obtain observational constrains on
the inflaton potential from a synthetic data-set.

\section{Projections for Planck: MCMC analysis}
\label{Projections for Planck: MCMC analysis}
In this section, we constrain the inflaton potential in the hypothetical case that Planck
detects tensors, with $r = 0.1$, but fails to positively detect any non-Gaussianities,
implying $|f_{NL}| \lesssim 5$.  Such an observation would certainly be consistent with
canonical inflation, for which $c_s = 1$.  However, it would also be consistent with a
host of more exotic, non-canonical models, such as k-inflation and DBI.  The speed of
sound varies in such models and gives rise to detectable non-Gaussianities if $c_s \ll
1$.  From Eq. (\ref{ng}), we have for the case of DBI inflation, 
\begin{equation}
\label{fnl}
f_{NL} = \frac{35}{108}\left(\frac{1}{c^2_s} - 1\right).
\end{equation}
However, even for non-detectable values of $f_{NL}$, we see that the sound speed can vary
considerably, $0.25 \lesssim c_s \leq 1$.  In non-canonical models, there is
another way to measure $c_s$, however,  independent of the strength of non-Gaussianities.
These models predict a modified consistency relation \cite{Garriga:1999vw},
\begin{equation}
r = -8c_s n_t,
\end{equation}
where $n_t$ is the tensor spectral index.  The downside is that this relation will be
very difficult to reliably measure any time in the near future.  Even a direct detection
of primordial gravity waves from proposed BBO \cite{BBO} or DECIGO \cite{Seto:2001qf} will only measure this relation
to within an accuracy of $50\%$ \cite{Smith:2006xf}.  The uncertainty in the value of $c_s$ thus remains.  How might
this affect our ability to reconstruct $V(\phi)$?

In models with a variable speed of sound, the scalar and tensor amplitudes are given during slow roll by,
\begin{equation}
\label{scal}
P_{\mathcal R}(k) = \frac{1}{8\pi^2 M^2_{\rm Pl}}\left.\frac{H^2}{c_s
\epsilon}\right|_{kc_s = aH},
\end{equation}
and
\begin{equation}
\label{tens}
P_h(k) = \frac{2}{\pi^2}\left.\frac{H^2}{M^2_{\rm Pl}}\right|_{k=aH},
\end{equation}
where we note that scalar perturbations freeze out upon exiting the sound horizon, whereas tensor perturbations do so
upon exiting the Hubble radius. 
Often in the literature, the tensor/scalar ratio is written $r = 16c_s \epsilon$,
obtained from Eqs. (\ref{scal}) and (\ref{tens}) by neglecting this difference in freeze out times
(assuming that $H$ is constant).  This is not strictly correct.  While this difference enters at
2$^{nd}$-order in slow roll, it can be important if $c_s$ is small\footnote{The
limited accuracy of this expression was recently noted in Ref. \cite{Agarwal:2008ah}, where numerical
spectrum calculations reveal discrepancies of up to 50\%.}.  The evolution of
$H$ is determined by
\begin{equation}
\frac{dH}{dN} = \epsilon H,
\end{equation}
where $N$ is defined as the number of e-folds before the end of inflation.
From the horizon crossing
conditions we form the ratio,
\begin{equation}
\frac{H_t}{H_s} = \frac{1}{c_s}\frac{a_s}{a_t},
\end{equation}
where the subscripts $t$ and $s$ indicate that these values are measured when the tensor and scalar perturbations exit their respective horizons.  
If we suppose that $\epsilon$ is constant between these crossing times (an excellent assumption during
slow roll), this ratio becomes
\begin{equation}
e^{\epsilon(N_t - N_s)} = c_s^{-1} e^{N_t-N_s},
\end{equation}
with the result that the change in the number of e-folds between the time that a certain scalar $k$-mode exits the sound horizon and the time the corresponding tensor mode exits the event horizon is given by,
\begin{equation}
N_s - N_t = \frac{{\rm ln}c_s}{\epsilon-1}.
\end{equation}
The tensor/scalar ratio is then\footnote{See Ref. \cite{Lorenz:2008et}
for an alternative derivation.}
\begin{eqnarray}
r = \frac{P_h(k)}{P_{\mathcal R}(k)} = 16\epsilon c_s \left(\frac{H_t}{H_s}\right)^2 &=&
16\epsilon c_s e^{2\epsilon(N_t - N_s)} \nonumber \\
&=& 16 \epsilon c_s^{\frac{1+\epsilon}{1-\epsilon}}.
\end{eqnarray}

The scalar spectral index of the power spectrum is also affected by the variable speed
of sound.  At lowest-order in slow roll,
\begin{equation}
n_s = 1 - 4\epsilon + 2\eta - 2s,
\end{equation}
where $\eta$ and $s$ are defined in Eq. (\ref{eq:flowparams}).
The tensor spectral index is written
\begin{equation}
n_t = -2\epsilon,
\end{equation}
which leads to the modified consistency relation at lowest order, $r =
-8c_sn_t$.

We are now in a position to determine what effect a non-detection of $f_{NL}$ has on
reconstruction efforts.  From our results in the previous section Eq. (\ref{V V' V''})
we can link $V(\phi)$ to
the observables $P_{\mathcal R}$, $r$, $n_s$, and $f_{NL}$.  In this section we work to lowest-order in
slow roll and assume that $c_s$ is constant.  We neglect any running of the
spectra and thus work to reconstruct $V(\phi_0)$ and its first two derivatives. In terms of observables these are written,
\begin{eqnarray}
\label{V V' V'' 2}
\frac{V(\phi_0)}{M_{\rm Pl}^4} &=& \frac{\pi^2}{2} P_{\mathcal R}r\left[3 - \frac{\gamma^2r}{8(\gamma +
1)}\right],\cr
\frac{V'(\phi_0)}{M_{\rm Pl}^3} &=& \frac{\sqrt{2}\pi^2}{16} P_{\mathcal R}r^{3/2}\gamma\left[6 -
\frac{\gamma}{2}\frac{(4n_s + r\gamma -4)}{\gamma + 1}\right],\cr
\frac{V''(\phi_0)}{M_{\rm Pl}^2} &=& \frac{\pi^2}{4} P_{\mathcal R}r\gamma\left[3\left(n_s - 1 +
\frac{3}{8}r\gamma\right) \right. \cr
&&\left.-\frac{\gamma}{2}\frac{\left(4n_s+r\gamma -
4\right)^2}{\gamma + 1}\right],
\end{eqnarray}
where the observables are to be measured at the scale corresponding to $\phi_0$.  We
have used $\gamma$ rather than the observable $f_{NL}$ in these expressions for simplicity.  It is now clear how any uncertainty in $\gamma$ will
translate into
uncertainties in the potential.  The height of the potential, $V(\phi_0)$, will be least
affected, however, $V'(\phi_0)$ and $V''(\phi_0)$ will be more strongly so because they
are directly proportional to $\gamma$.  

In order to rigorously analyze the effects of an
unresolved speed of sound, we constrain $V(\phi_0)$ and its first two derivatives using
Markov Chain Monte Carlo (MCMC) and a simulated, Planck-precision CMB data-set.   
We use the publicly available
\texttt{CosmoMC}\footnote{http://cosmologist.info/cosmomc/}  \cite{Lewis:2002ah} package to explore the
inflationary parameter space.   We constrain $V(\phi)$ arising from non-canonical models
by allowing $c_s$ to vary in the analysis, and compare this case to constraints obtained
on $V(\phi)$ arising from canonical inflation, obtained by fixing $c_s = 1$. 

MCMC techniques \cite{Christensen:2000ji,Christensen:2001gj,Verde:2003ey,Mackay} sample the likelihood surface of model parameters,
\(\mathcal{L}({\bf d}|{\bm \theta})\),
where \({\bf d}\) represents the n-dimensional data and \({\bm \theta}\) the n-dimensional parameter
vector.
The likelihood function
relates any prior knowledge of the parameter values, \(\pi({\bm \theta})\), to the posterior
probability distribution via Bayes' theorem,
\begin{equation}
\label{bayes}
p({\bm \theta}|{\bf d}) = \frac{\mathcal{L}({\bf d}|{\bm \theta})\pi({\bm \theta})}{P({\bf d})} =
\frac{\mathcal{L}({\bf d}|{\bm \theta})\pi({\bm \theta})}{\int
\mathcal{L}({\bf d}|{\bm \theta})\pi({\bm \theta})d{\bm \theta}}.
\end{equation}
The posterior probability distribution function (PDF) of a single parameter \(\theta_i\) can then be
obtained by marginalizing \(p({\bm \theta}|{\bf d})\) over the remaining parameters,
\begin{equation}
p(\theta_i|{\bf d}) = \int p({\bm \theta}|{\bf d})d\theta_1 \cdots d\theta_{i-1} d\theta_{i+1}\cdots d\theta_{n-1}.
\end{equation}
\begin{figure*}
\label{figure1}
\includegraphics[width=4.1in]{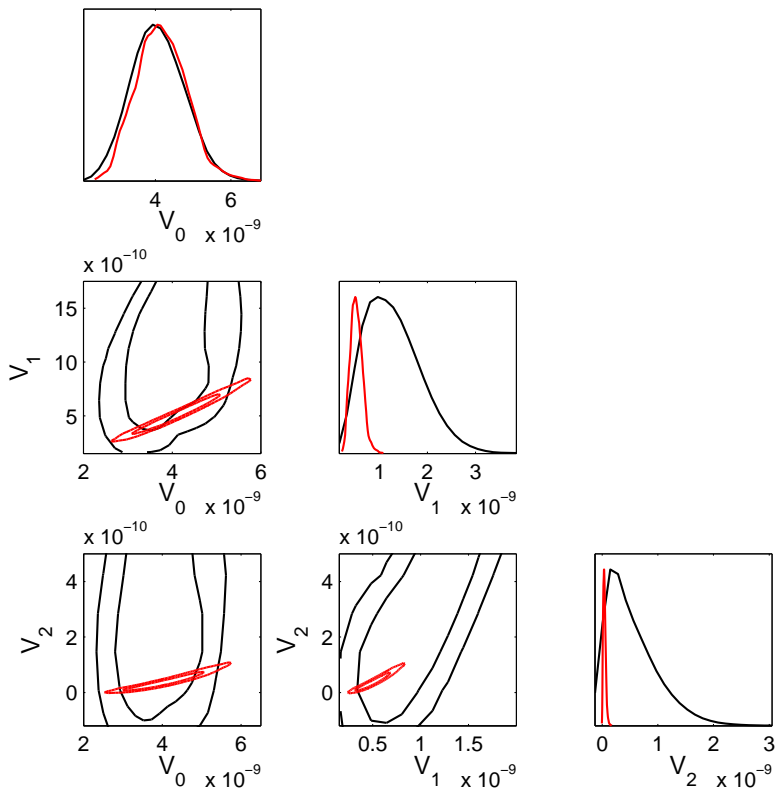}
\caption{Marginalized 1- and 2-D PDF's of the potential parameters, $V_0 = V(\phi_0)$, $V_1 = V'(\phi_0)$, and $V_2 = V''(\phi_0)$ in units of $M_{\rm Pl}$.  
The small red contours correspond to 1- and 2-$\sigma$ errors in canonical inflation, in which $c_s$ is held fixed at 1.
The substantially larger black contours correspond to non-canonical inflation in which
$c_s$ is allowed to vary in the chains between 0.25 and 1.}
\end{figure*}
In this study, we only constrain the inflationary parameters ($V(\phi)$, $V'(\phi)$,
$V''(\phi)$, and $c_s$), fixing the late-time parameters at the values: baryon and CDM
densities, $\Omega_bh^2 = 0.022$ and $\Omega_ch^2 = 0.104$, the angular diameter
distance at decoupling, $\theta_s = 1.04$, and the optical depth to reionization, $\tau
= 0.093$.  Degeneracies amongst the late-time parameters and the power spectrum do
exist, in particular, $n_s$ is still moderately degenerate with $\Omega_bh^2$.  However,
our goal in this study is to compare the constraints obtained on $V(\phi)$ when $c_s$ is
allowed to vary and when it is held constant at $c_s = 1$.  In both of these cases the
spectrum is completely described by the parameters $P_{\mathcal R}$, $r$, $n_s$, and
$n_t$, and so the introduction of $c_s$ as a free parameter will not introduce new
degeneracies.  

Our Planck-precision simulation comprises temperature, E- and
B-mode polarization data out to $\ell = 2500$ with a sky coverage of $65\%$. For simplicity, we
consider a single-channel experiment in which the full width at half maximum of the
Gaussian beam, $\theta_{\rm fwhm} = 7'$, and the root mean square of the pixel noise,
$\sigma^2_T = \sigma^2_P/2 = 30.25 \mu {\rm K}^2$.  As a fiducial model we choose a
power-law spectrum with $P_{\mathcal R}(k = 0.002 h\,{\rm Mpc}^{-1}) = 2.3 \times
10^{-9}$, $n_s = 0.97$, and $r=0.1$. We assume the inflationary consistency relation for the tensor spectral
index, $n_t = -\gamma r/8$.  We impose the prior $c_s \in
[0.25,1]$, consistent with undetectable non-Gaussianities.  
We assume purely adiabatic initial perturbations and spatial flatness, and adopt a top-hat prior on the age of the
universe: \(10 < t_0 < 20 \, {\rm Gyr}\).  We utilize the Metropolis-Hastings sampling
algorithm and measure convergence across 8 chains with the Gelman-Rubin R statistic.  

We present our results in Figure 1 and Table 1. 
Both best-fit reconstructions fit the data equally well: $-2{\rm
ln}\mathcal{L} = 7931.29$ for the canonical and $-2{\rm ln}\mathcal{L} =
7931.16$ for the non-canonical reconstruction.  The two models are thus statistically
degenerate.  
The red contours in the figure
correspond to 1- and 2-$\sigma$ errors for
canonical inflation, for which $c_s$ is held fixed at 1.
The black contours correspond to non-canonical inflation in which
$c_s$ is allowed to vary in the chains between 0.25 and 1.  Indeed, the constraints on $V(\phi)$
worsen significantly if $c_s$ is included in the reconstruction.  
This is a significant impediment to the program of potential reconstruction: if we allow for the possibility of an arbitrary speed of sound, reconstruction of the potential is severely degraded in the absence of a measurement of the amplitude of non-Gaussianities.
Table I indicates that the marginalized errors on $V_1$ and $V_2$
increase by an order of magnitude in going from canonical to
non-canonical inflation.  While we have
used Eq. (\ref{fnl}) specific to DBI inflation, the fact that a failure to resolve non-Gaussianities
precludes a unique inversion of the power spectrum is a general result.  We therefore expect this problem
to persist in general non-canonical models, however, the degree of the effect is model dependent.

It is interesting to note how well constrained the canonical model is.  In
particular, the absolute errors on $V''$ are {\it smaller} than those obtained on $V'$.  The
dominant term in the error on $V''$ is proportional to $3\pi^2 \delta r P_{\mathcal R}(n_s
-1)/2$,
whereas the dominant term in the error on $V'$ is $\delta V' \sim P_{\mathcal R}(r + \delta r)^{3/2}
- V'$.  For Planck-precision errors on $r$, the two terms in $\delta V'$ are of the same
order, and so $\mathcal{O}(\delta V') \sim \mathcal{O}(V')$.  For the values of our
fiducial model, $(r + \delta r)^{3/2} \approx \delta r$, and so roughly we have
\begin{equation}
\delta V'' \approx \frac{3}{2}\pi^2(n_s -1)\delta V' \sim \mathcal{O}(10^{-1})\delta
V'.
\end{equation}

In Figure 2 we plot the reconstructed best fit potentials for each model.  As indicated
by the error contours and by this figure, the speed of sound has the effect of rendering
steep potentials viable.  This is precisely the novel aspect that initially popularized DBI
inflation, but as we see here, is a bane to reconstruction efforts if $f_{NL}$ goes
undetected.  However, the detection of tensor modes still reliably determines the energy
scale of inflation.  For $r=0.1$, the uncertainty introduced by $c_s$ is only around $1\%$. 
The energy scale reconstructed from the observables is measured at $\phi = 0$ in 
Figure 2.

\begin{table}
\begin{tabular}{||l|c|c|c||}
\hline
Model & $V_0 \times 10^9 M_{\rm Pl}^{-4}$ & $V_1\times 10^{10} M_{\rm
Pl}^{-3}$
& $V_2\times 10^{11} M_{\rm Pl}^{-2}$ \\
\hline
canonical & $3.7^{+1.9}_{-0.7}$ & $4.3^{+3.7}_{-1.2}$ &
$2.5^{+7.2}_{-1.8}$ \\
\hline
non-canonical & $3.6^{+1.9}_{-0.9}$ & $16^{+10}_{-12}$ & $91^{+80}_{-90}$\\
\hline
\end{tabular}
\caption{Marginalized 2-$\sigma$ errors on the potential parameters of
each model.}  
\end{table}

\section{Projections for Planck: Flow Analysis}

While the MCMC method utilized in the last section is a rigorous Bayesian analysis of
the parameter space, it performs poorly if too many free parameters are included in the
fit. This is especially true if some of these parameters are degenerate or poorly constrained by the
data.  It is therefore not feasible to go much beyond the lowest-order analysis ($c_s =$
const.) conducted in the previous section.   
In this section, we use the flow formalism \cite{Hoffman:2000ue,Kinney:2002qn} as a
method for reconstructing inflationary potentials compatible with future observations.
While not statistically rigorous, the flow method is well suited to studying the
inflationary parameter space to arbitrarily high order in slow roll.   It is also
capable of reconstructing models with a varying sound speed.   In this
section, we utilize the flow formalism in order to study parameter
constraints in the event that a tensor signal is detected in the future, but non-Gaussianities
are not.  This is the same scenario considered in the last section, but here
we examine the effects of allowing a time-varying speed of sound.  We
conclude this section by considering the case of a positive detection of non-Gaussianities. 
\begin{figure}
\centerline{\includegraphics[angle=270,width=3.75in]{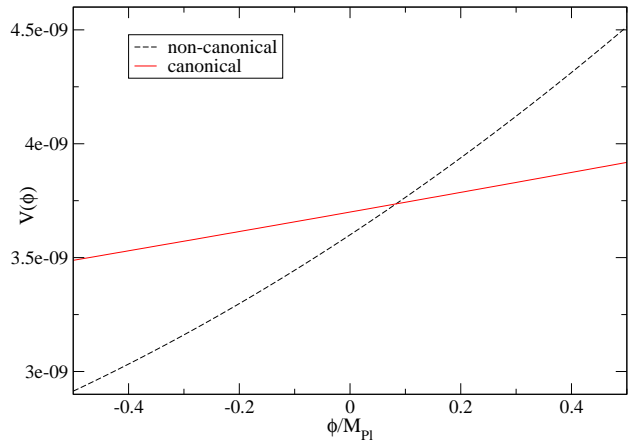}}
\caption{Potentials reconstructed from the best-fit parameters of Table
1.  Note the increased steepness of the non-canonical potential (black
dashed) relative to the canonical potential (red solid).} 
\end{figure}

The flow formalism is based on the property of the flow parameters defined in Eq. (\ref{eq:flowparams}), that their evolution in terms of the
number of e-folds $N$ can be described by a set of first order differential equations. Taking successive derivatives of the flow parameters with
respect to $N$, we obtain the {\it flow equations}
\begin{eqnarray}
\label{eq:flowequations}
\epsilon &=& \frac{1}{H}\frac{d H}{d N},\cr
\frac{d \epsilon}{d N} &=& \epsilon\left(2 \eta - 2 \epsilon - s\right),\cr
\frac{d \eta}{d N} &=& -\eta\left(\epsilon + s\right) + {}^2 \lambda,\cr &\vdots& \cr
\frac{d {}^\ell \lambda}{d N} &=& -{}^\ell \lambda \left[\ell \left(s + \epsilon\right) - \left(\ell - 1\right) \eta\right] + {}^{\ell + 1}\lambda,\cr
s &=& \frac{1}{\gamma}\frac{d \gamma}{d N},\cr
\frac{d s}{d N} &=& -s \left(2 s + \epsilon - \eta\right) + \epsilon \rho,\cr
\frac{d \rho}{d N} &=& -2 \rho s + {}^2 \alpha,\cr &\vdots& \cr
\frac{d {}^{\ell} \alpha}{d N} &=& -{}^{\ell} \alpha \left[\left(\ell + 1\right) s + \left(\ell - 1\right) (\epsilon-\eta)\right] + {}^{{\ell} + 1} \alpha,
\end{eqnarray}
where the derivative of a flow parameter is related to terms of higher-order in the flow hierarchy. In practice, the flow equations are truncated at some finite order $\mathcal{M}$ and $\mathcal{N}$ by requiring that $^{\ell}\lambda =\,\, ^{\ell}\alpha=0$ for all $\ell \geq \mathcal{M}+1$ and $\mathcal{N}+1$, respectively, at the initial time. By Eq. (\ref{eq:flowequations}), these remain zero for all time with the result that the trajectory so obtained is an {\it exact} solution to the inflationary equations of motion. 

This truncated system of differential equations can be solved numerically by specifying the set of initial conditions for the flow parameters
$\left. \left[H,\epsilon, \dots, ^{\mathcal{M}}\lambda;\gamma,s,\dots,{}^{\mathcal{N}}\alpha\right]\right|_{N_{0}}$ at some time $N_{0}$. The
solution will then be a particular inflationary path $\left[H(N), \epsilon(N), \dots,
^{\mathcal{M}}\lambda(N);\gamma(N),s(N),\dots,{}^{\mathcal{N}}\alpha(N)\right]$ in the $\mathcal{(M+N)}$-dimensional parameter space. 
In fact, the expansions Eqs. (\ref{eq:polynomial H}) and
(\ref{eq:polynomial gamma}) yield exact analytic solutions of the flow equations,
where the expansions can be taken about any point along the inflationary trajectory
\cite{Liddle:2003py}. 
In this paper,
we truncate the flow hierarchy to second order in the $H$-tower, and to first order in the $\gamma$-tower so that
\begin{equation}
\label{eq:paper truncation}
{}^{\ell}\lambda = {}^{\ell'}\alpha=0,
\end{equation}
for all $\ell>2$ and $\ell'>1$. The resulting solution to the flow equations will then
be a specific path $\left[\epsilon(N), \eta(N),\xi(N),s(N),\rho(N)\right]$ in the
5-dimensional parameter space. The inclusion of $s$ and $\rho$ allows us to
generate models with a varying speed of sound.  
We solve the flow equations from the time the quadrupole leaves the horizon at $N_{\rm quad}$ up until the smallest scale modes detectable in the CMB are generated, a period
typically spanning 10 e-folds.  We restrict this analysis to models that do not
violate causality,  $c_{s}\leq1$. We then calculate the observables from the
values of the flow parameters at $N=N_{\rm quad}$, 
\begin{eqnarray}
\label{eq:observs}
n_{s}&=&1-4\epsilon+2\eta-2s-2(1+C)\epsilon^2-(3+C)s^2\cr &&-\frac{1}{2}(3-5C)\epsilon\eta-\frac{1}{2}(11+3C)\epsilon s+(1+C)\eta s\cr&& +\frac{1}{2}(1+C)\epsilon \rho +\frac{1}{2}(3-C) ({}^2 \lambda), \cr r&=&16c_{s}\epsilon\left[1+\frac{1}{2}(\epsilon-\eta)(C-3)+\frac{s}{2}(C+1)\right], \cr \alpha &=& -\left(\frac{1}{1-\epsilon-s}\right)\frac{dn_{s}}{dN},\cr f_{NL}&=&\frac{35}{108}\left(\frac{1}{c_s^2}-1\right),
\end{eqnarray}
where
\begin{equation}
\label{eq:C}
C=4({\rm ln}2+\gamma)-5,
\end{equation}
and $\gamma \simeq 0.577$ is the Euler-Mascheroni constant, which should not
be confused with the inverse sound speed Eq. (\ref{eq:gamma}).

\subsection{Detection of \boldmath{$r$} and no detection of \boldmath{$f_{NL}$}}

\begin{figure*}
\label{Fig:figure3}
\includegraphics[width=7.0in]{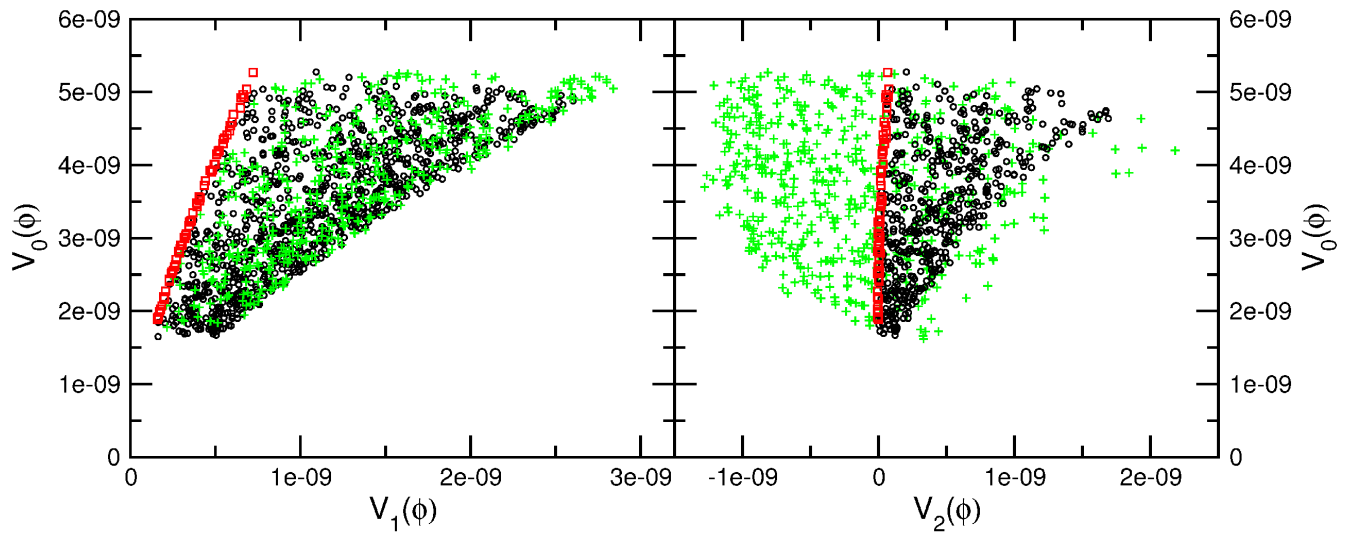}
\caption{Results of the flow analysis for the case of detection of $r$, with $r=0.1\pm0.05$ but no $f_{NL}$ ($f_{NL}<5$).  Red squares are canonical models, black circles DBI inflation with constant speed of sound, and green crosses DBI models with variable speed of sound. The case of variable speed of sound opens up the parameter space for which $V''_{0}<0.$}
\end{figure*}
We are interested in collecting
models compatible with the simulated data of Section III, 
\begin{eqnarray}
\label{eq:accept abservs}
n_{s}&=&0.97\pm 0.0036, \cr r&=&0.1\pm 0.05, \cr \alpha&=&0.0\pm 0.005, \cr 10^9P_{\mathcal{R}}&=&2.3\pm 0.102,
\end{eqnarray}
where the errors are 1-$\sigma$ projections for Planck, quoted at $k = 0.002\,h{\rm
Mpc}^{-1}$. 
The observables calculated at the quadrupole by the flow method are easily
extrapolated to  $k = 0.002\,h{\rm Mpc}^{-1}$ because the spectra are power-laws.
In order to efficiently generate models with power-law spectra,  we truncate the $H$-tower at $\xi^2$.  The inclusion of
higher-order terms not only affects the running, but may even make a non-negligible contribution
to higher-order $k$-dependencies in the power spectrum.  Since we do not have analytic
expressions\footnote{This issue is avoided by calculating the power spectra
numerically, as done in \cite{Agarwal:2008ah}.} for these higher-order terms, we can not be sure that they are under control. By working only up to order $\xi^2$, and by ensuring that the running is small, we can be
certain that all higher-order terms will likewise be small, since they
are at least of
order $\xi^4$.  Most viable models at this order naturally have small running anyway, since
large negative running at order $\xi^2$ typically leads to an insufficient amount of inflation
\cite{Easther:2006tv}. 

We now use this methodology to stochastically generate models of inflation consistent with the constraints Eq.
(\ref{eq:accept abservs}).  We are interested in models which fail to generate observable non-Gaussianities,
corresponding to a speed of sound in the range $c_s \in [0.25,1]$ (c.f. Eq. (\ref{fnl})).  We first seek to reconstruct models with a constant speed of sound in order to make contact with the results of
section 3, and so the $\gamma$-tower is simply replaced by $c_s =$ const.
After obtaining results for constant speed of sound, we  perform the analysis for a variable speed of sound by
allowing $s$ and $\rho$ to vary. 

We present our results in Figure 3.  
We plot the values of $V_0$, $V_1 = V'_0$, and $V_2 = V''_0$ for canonical inflation, where $c_s = 1$ (red squares), DBI
inflation with constant speed of sound (black circles), and DBI inflation with varying speed of sound (green
crosses) using Eqs. (\ref{V V' V''}). It can easily be seen that, even
though we can not make any rigorous statistical arguments using the flow formalism, we successfully reproduce the results found in the previous section using MCMC techniques. 
The most striking consequence of a variable speed of sound is that a large region of parameter space for
which $V''_0 < 0$ opens up.  To understand this, note from Eq. (\ref{V V' V''}) that the sign of $V''$ to
lowest-order is determined by the sign and magnitude of $\eta$.  In the case of constant sound speed, the
value of $\eta$ is tightly constrained by the spectral index, $n_s = 1 - 4\epsilon + 2\eta$.  However, if
$c_s$ is allowed to vary, then the spectral index becomes
\begin{equation}
n_s = 1 -4\epsilon + 2\eta -2s,
\end{equation}
to lowest-order.  For a given value of $n_s$, $\eta$ can be taken more negative by
suitably adjusting $s$.  From Eq. (\ref{V V' V''}), this allows a wide range of
parameter space for which $V''_0 < 0$ to be brought into agreement with the data.
Physically, the steepening of the potential that results from tuning
$\eta$ is mitigated by the increased warping of the geometry, which
is related to the value of the flow parameter $s$.

\begin{figure*}
\label{figure4}
\includegraphics[width=7.0in]{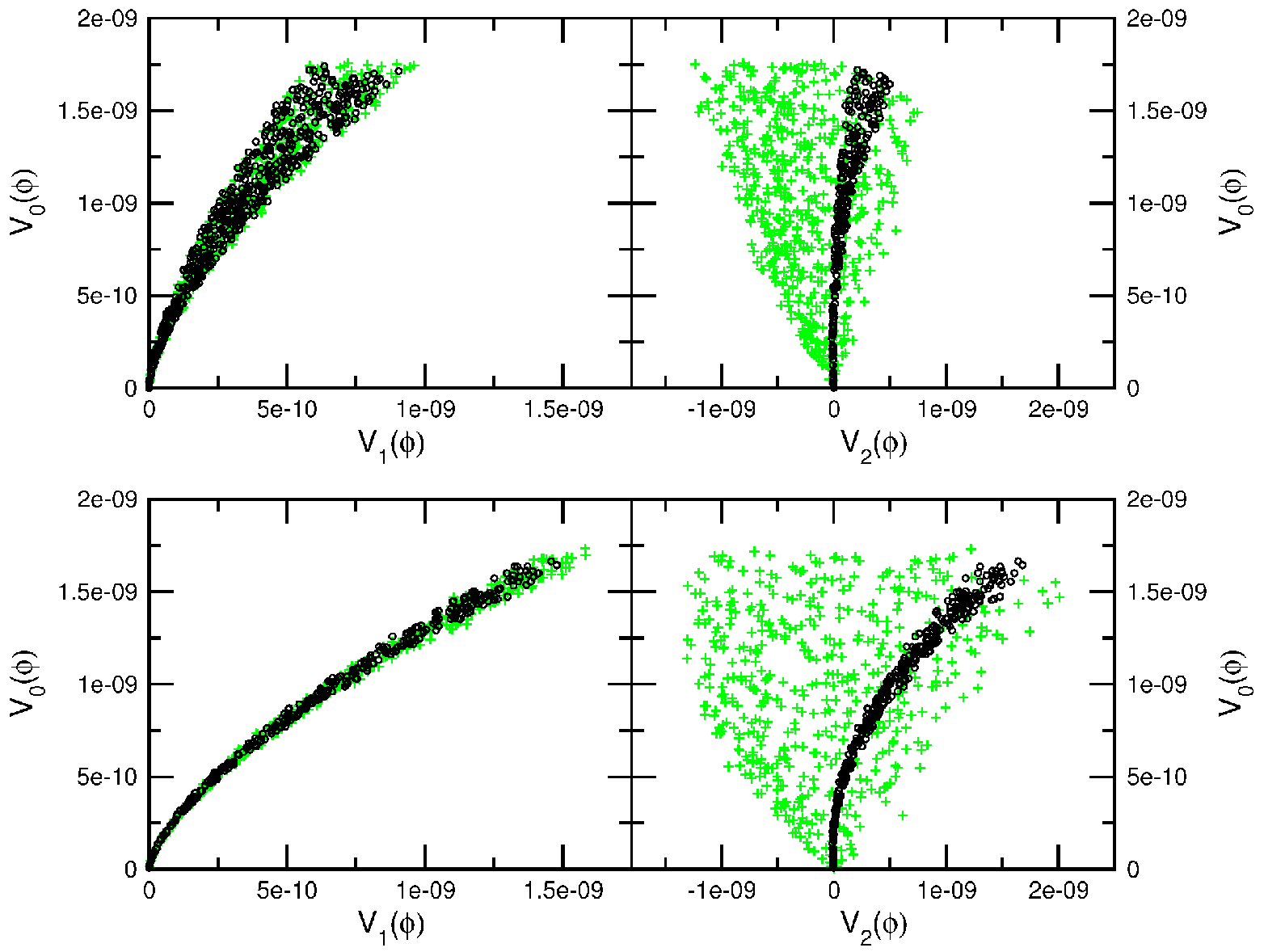}
\caption{Results of the flow analysis for the case of detection of $f_{NL}$ but no $r$ ($r \leq 0.05$). The top figure corresponds to models with $f_{NL}=10 \pm 5$, and the bottom one to models with $f_{NL}=40 \pm 5$.  Black circles are DBI models with constant speed of sound, and green crosses DBI models with variable speed of sound.}
\end{figure*}

\begin{figure*}
\label{figure5}
\includegraphics[width=7.0in]{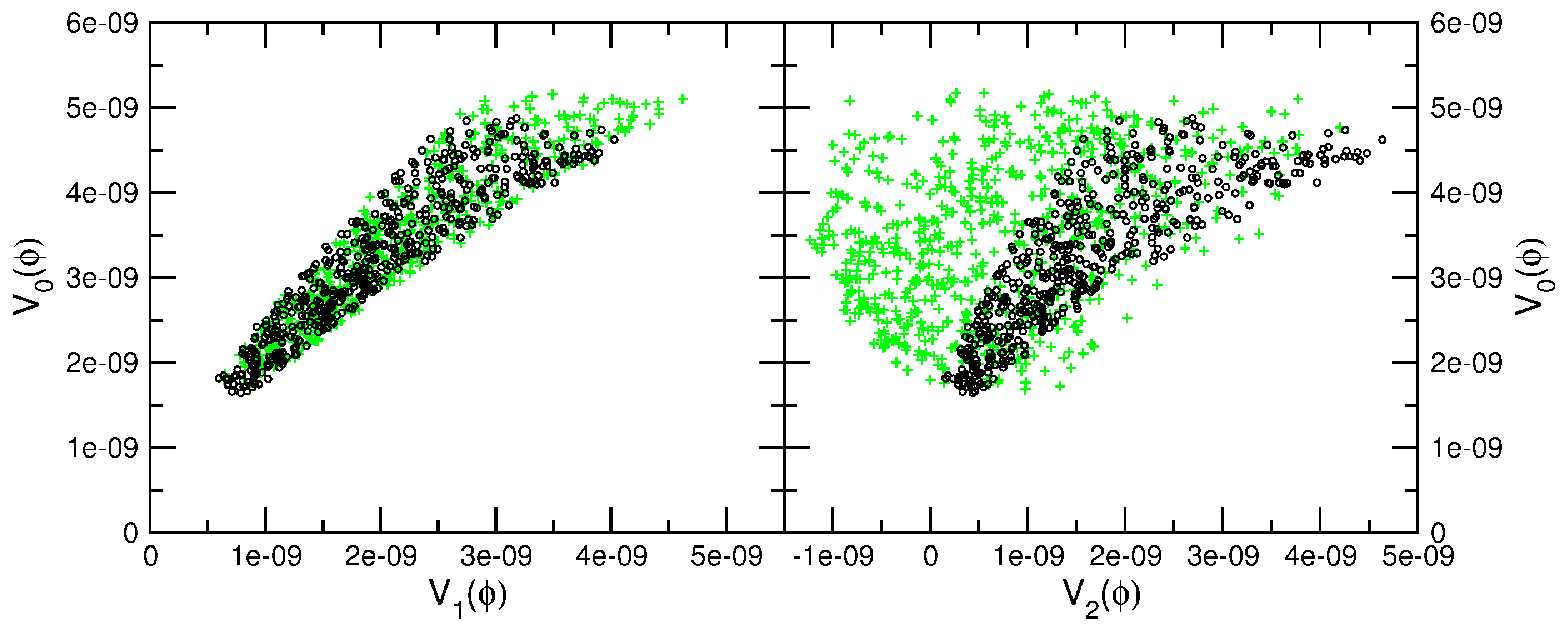}
\caption{Results of the flow analysis for the case of detection of $r$, with $r=0.1\pm0.05$ and $f_{NL}$, with $f_{NL}=10\pm5$. Black circles are DBI models with constant speed of sound, and green crosses DBI models with variable speed of sound.}
\end{figure*}

\subsection{Detection of \boldmath{$f_{NL}$}}

In this section, we are interested in models which are characterized by
detectable non-Gaussianity for both observational outcomes of gravity waves. We start by considering the case in which future data fails to detect any gravity waves, but
{\it does} successfully detect evidence for non-Gaussian fluctuations.  
While it is well known that a failure to detect tensors precludes a lower bound on the energy scale of
inflation, it might be hoped that a positive detection of non-Gaussianities yields improved constraints
relative to the worst case scenario in which neither is detected.  
The simulated data used thus far in
the study does not apply to this case.  However, taking the scalar perturbations to be the same, we collect
models that satisfy Eq. (\ref{eq:accept abservs}) for $n_s$, $P_{\mathcal
R}$, and  $\alpha$.  In order for Planck to
fail to detect tensors, we suppose that the null hypothesis ($r = 0$) is verified for $r \leq 0.05$.

We show the results of the flow analysis in Figure 4.  We consider detection of non-Gaussianities at two
different levels:  $f_{NL} = 10$ and $f_{NL} = 40$, with 1-$\sigma$ errors $\Delta f_{NL} = \pm 5$.  By failing to detect $r$, we only succeed in
placing an upper bound on the values of $V_0$ and $V_0'$. However, a determination of $\gamma$
tightly correlates $V_0$ with both $V_0'$ and $V_0''$.  From Eq. (41) we have that
\begin{equation}
V \sim \left(\frac{P_{\mathcal{R}}(k)}{\gamma^2}\right)^{1/3}V'^{2/3},
\end{equation}
and 
\begin{equation}
V \sim \frac{P_{\mathcal{R}}(k)}{\gamma}\left[-1 +
\sqrt{\frac{4V''}{3P_{\mathcal{R}}(k)}\frac{1}{n_s-1}}\right].
\end{equation}
The uncertainty in $r$ only allows us to constrain the above relationships between the potential
parameters.  These relations give the distinctive band structure seen in Figure 4.  The effect of
the detection of non-Gaussianities on these constraints is two-fold: the value of $\gamma$
controls the `slope' of the bands, and, as we show next, the relative error between $V$-$V'$ and
$V$-$V''$.

The contribution of the error $\Delta f_{NL}$ to $V'$
and $V''$ depends on the absolute value of $f_{NL}$.  Since this is the dominant source of error, a
determination of $f_{NL}$ can thus improve constraints.  To see this, consider the case of $V'$.  Including only
the contribution of $f_{NL}$ to the error on $V'$, we have
\begin{equation}
\delta V' \propto \left(\sqrt{f_{NL} + \Delta f_{NL}} - \sqrt{f_{NL} - \Delta f_{NL}}\right)\delta V,
\end{equation}
and expanding the square roots for small $\Delta f_{NL}$ relative to $f_{NL}$ gives
\begin{equation}
\delta V' \propto \frac{\Delta f_{NL}}{\sqrt{f_{NL}}}\delta V.
\end{equation}
Thus, a determination of $f_{NL}$ correlates the errors in $V$ and $V'$, and similarly for $V$ and $V''$
(c.f. Figure 4).
Additionally, the larger the value of $f_{NL}$, the smaller this relative error between the potential
coefficients.  
Therefore, while a detection of $r$ allows us to constrain the absolute scale of the inflationary potential and its derivatives, a detection of $f_{NL}$ gives us information about the {\it relative} values of these parameters, as well as being a smoking gun for exotic physics.

Finally, we consider the case where {\it both} $r$ and $f_{NL}$ are
detected. We collect models that satisfy Eq. (\ref{eq:accept abservs}),
and produce detectable non-Gaussianities at the level of $f_{NL}=10$ with
1-$\sigma$ errors $\Delta f_{NL}=\pm5$. We present the results of the
flow analysis in Figure 5, where it can be seen that we again
obtain the band-shaped distributions for the potential parameters.
Following the reasoning of the previous sections, we see that the
detection of $r$ places bounds on the potential parametes, where the
detection of non-Gaussianities determines the relative error amongst
them.

\section{Conclusions}
We have considered potential reconstruction forecasts for the upcoming Planck Surveyor CMB mission in the
context of non-canonical inflation.  We
have focused on the following observational outcomes: a detection of a tensor signal and a null detection
of non-Gaussianities, a null detection of tensors and a positive
detection of non-Gaussianities, and a positive detection of both tensors
and non-Gaussianities.  To
explore the first case, we perform a Markov Chain Monte Carlo analysis on a simulated Planck-precision
data-set.  We obtain constraints on the potential parameters $V_0$, $V_0'$, and $V_0''$, and the sound
speed $c_s$.  A null detection of non-Gaussianities at Planck resolution corresponds to $|f_{NL}| \lesssim
5$, with the result that $c_s$ is only constrained to lie within the range $[0.25,1]$ for the case of DBI
inflation.  Since $c_s$ also affects the form of the power spectrum, a failure to constrain it precludes
us from inverting our spectrum observation to obtain the inflaton potential.  The errors on $V_0'$ and
$V_0''$ increase by an order of magnitude relative to constraints obtained on canonical inflation.  

We next used the flow formalism to investigate the effect of allowing the speed of sound to vary over the
course of inflation.  This requires the introduction of terms governing its evolution.  Unlike MCMC
methods, the flow formalism is well suited to studying reconstructions to high-order.  The strongest
effect of a variable speed of sound is a much increased error on the second derivative of $V(\phi)$.  A
variable speed is generic to models such as DBI inflation.  The worsening
of the constraint on $V_0''$ results from the fact that there is additional freedom in tuning the
geometry in order to get the same observables. (We note that this has been a very conservative analysis as far as our constraints on $f_{NL}$ are
concerned.  If one considers $f_{NL}^{\rm equil}$, the non-Gaussianities associated with equilateral
triangles in $k$-space, the Planck detection threshold could be as large as $f_{NL}^{\rm equil} \lesssim
60$ \cite{Liguori:2008vf}.) 

We also used the flow approach to investigate the
observational outcome of a detection of different
levels of non-Gaussianity. We first consider the case of a null measurement of $r$. Because neither $V_0$ or $V_0'$ are bounded
from below, and since a larger level of non-Gaussianity is generated by
larger values of $V_0'$, we find that the errors on the individual
parameters $V_0$, $V_0'$, and $V_0''$ grow relative to the case where
neither are detected. However, these errors are strongly correlated, and
it is nonetheless possible to tightly constrain combinations of these
parameters. The parameters $r$ and $f_{NL}$ therefore play complementary
roles in potential reconstruction. Finally, we consider the case where
{\it both} $r$ and $f_{NL}$ are detected. The complementary role that $r$
and $f_{NL}$ play in potential reconstruction is again clear, where gravity waves place bounds on $V_{0}$ and $V'_{0}$ from above and below, and non-Gaussianities determine the width of the bands in the $V$-$V'$ and $V$-$V''$ planes.

The approach taken in this paper has been phenomenological.  We have treated the potential and $c_s$ as
freely tunable parameters. 
 We have
purposefully kept the study general to encompass the larger class of $k$-inflation and other
non-canonical models based on effective field theory.
However, it is very likely that many of the potentials and sound speeds sampled in
this analysis do not represent realistic compactification scenarios in string theory, and might not
correspond to realizations of DBI inflation. What we have shown
in this study is that, if the observational outcomes considered here are reflected in the actual data, then
our ability to reconstruct the potential is significantly weakened. While we ultimately hope that the best-case scenario will be realized with a detection of both tensors and non-Gaussianities, in the meantime we can only wait with guarded anticipation as to what Planck will reveal about the universe.   

\section*{Acknowledgments}
We acknowledge the use of the UB Physics Graduate Computing Facility and
thank Mark Kimball for assistance.
This research is supported in part by the National Science Foundation under grants
NSF-PHY-0456777 and NSF-PHY-0757693.  This work was supported by World Premier International
Research Center Initiative (WPI Initiative), MEXT, Japan.

\end{document}